# DBToaster: Higher-order Delta Processing for Dynamic, Frequently Fresh Views[*]

Yanif Ahmad  
yanif@jhu.edu  
Johns Hopkins University

Oliver Kennedy, Christoph Koch, and Milos Nikolic  
{christoph.koch, oliver.kennedy, milos.nikolic}@epfl.ch  
École Polytechnique Fédérale de Lausanne

## ABSTRACT

Applications ranging from algorithmic trading to scientific data analysis require realtime analytics based on views over databases that change at very high rates. Such views have to be kept fresh at low maintenance cost and latencies. At the same time, these views have to support classical SQL, rather than window semantics, to enable applications that combine current with aged or historical data.

In this paper, we present *viewlet transforms*, a recursive finite differencing technique applied to queries. The viewlet transform materializes a query and a set of its higher-order deltas as views. These views support each other's incremental maintenance, leading to a reduced overall view maintenance cost. The viewlet transform of a query admits efficient evaluation, the elimination of certain expensive query operations, and aggressive parallelization. We develop viewlet transforms into a workable query execution technique, present a heuristic and cost-based optimization framework, and report on experiments with a prototype dynamic data management system that combines viewlet transforms with an optimizing compilation technique. The system supports tens of thousands of complete view refreshes a second for a wide range of queries.

## 1. INTRODUCTION

Data analytics has been dominated by after-the-fact exploration in classical data warehouses for multiple decades. This is now beginning to change: Today, businesses, engineers and scientists are increasingly placing data analytics engines earlier in their workflows to react to signals in fresh data. These dynamic datasets exhibit a wide range of update rates, volumes, anomalies and trends. Responsive analytics is an essential component of computing in finance, telecommunications, intelligence, and critical infrastructure management, and is gaining adoption in operations, logistics, scientific computing, and web and social media analysis.

Developing suitable analytics engines remains challenging. The combination of frequent updates, long-running queries and a large stateful working set precludes the exclusive use of OLAP, OLTP, or stream processors. Furthermore query requirements on updates often do not fall singularly into the functionality and semantics provided by the available technologies, from CEP engines to triggers, active databases, and database views.

Our work on *dynamic data management systems* (DDMS) in the DBToaster project [3, 19, 18] studies the foundations, algorithms and architectures of data management tools designed for large datasets that evolve rapidly through high-rate update streams. A DDMS focuses on long-running queries as the norm, alongside sporadic exploratory querying. A guiding design principle for DDMS is to take full advantage of incremental processing techniques that maximally reuse prior work. Incremental computation is central to both stream processing to minimize work as windows slide, and to database views with incremental view maintenance (IVM). DDMS aim to combine some of the advantages of DBMS (expressive queries over both recent and historical data, without the restrictions of window semantics) and CEP engines (low latency and high view refresh rates).

An example use case is algorithmic trading. Here, strategy designers want to use analytics expressible in a declarative language like SQL on order book data in their algorithms. Order books consist of the orders waiting to be executed at a stock exchange and change very frequently. However, some orders may stay in the order book relatively long before they are executed or revoked, precluding the use of stream engines with window semantics. Applications such as scientific simulations and intelligence analysis also exhibit entities which capture our attention for widely ranging periods of time, resulting in large stateful and dynamic computation.

The technical focus of this paper is on an extreme form of incremental view maintenance that we call *higher-order* IVM. We make use of discrete forward differences (delta queries) recursively, on multiple levels of derivation. That is, we use delta queries ("first-order deltas") to incrementally maintain the view of the input query, then materialize the delta queries as views too, maintain these views using delta queries to the delta queries ("second-order deltas"), and continue alternating between materializing views and deriving higher-order delta queries for maintenance. The technique's use of higher-order deltas is quite different from earlier work on trading off choices in which query *subexpressions* to materialize and incrementally maintain for best performance [27]. Instead, our technique for constructing higher-order delta views is somewhat reminiscent of discrete wavelets and numerical differentiation methods, and we use a superficial analogy to the Haar wavelet transform as motivation for calling the base technique a *viewlet transform*.

**Example 1** Consider a query $Q$ that counts the number of tuples in the product of relations $R$ and $S$. For now, we only want to maintain the view of $Q$ under insertions.

---

[*]This work was supported by ERC Grant 279804.

Denote by $\Delta_R$ (resp. $\Delta_S$) the change to a view as one tuple is inserted into $R$ (resp., $S$). Suppose we simultaneously materialize the four views

- $Q$ (0-th order),
- $\Delta_R Q = \text{count}(S)$ and $\Delta_S Q = \text{count}(R)$ (first order), and
- $\Delta_R(\Delta_S Q) = \Delta_S(\Delta_R Q) = 1$ (second order, a "delta of a delta query").

Then we can simultaneously maintain all these views using each other, using exclusively summation and avoiding the computation of any products. The fourth view is constant and does not depend on the database. Each of the other views is refreshed when a tuple is inserted by adding the appropriate delta view. For instance, as a tuple is inserted into $R$, we add $\Delta_R Q$ to $Q$ and $\Delta_R \Delta_S Q$ to $\Delta_S Q$. (No change is required to $\Delta_R Q$, since $\Delta_R \Delta_R Q = 0$.) Suppose $R$ contains 2 tuples and $S$ contains 3 tuples. If we add a tuple to $S$, we increment $Q$ by 2 ($\Delta_S Q$) to obtain 8 and $\Delta_R Q$ by 1 ($\Delta_S \Delta_R Q$) to get 4. If we subsequently insert a tuple into $R$, we increment $Q$ by 4 ($\Delta_R Q$) to 12 and $\Delta_S Q$ by 1 to 3. Two further inserts of $S$ tuples yield the following sequence of states of the materialized views:

| time point | insert into | $\|R\|$ | $\|S\|$ | $Q$ | $\Delta_R Q$ | $\Delta_S Q$ | $\Delta_R \Delta_S Q$, $\Delta_S \Delta_R Q$ |
|---|---|---|---|---|---|---|---|
| 0 | – | 2 | 3 | 6 | 3 | 2 | 1 |
| 1 | S | 2 | 4 | 8 | 4 | 2 | 1 |
| 2 | R | 3 | 4 | 12 | 4 | 3 | 1 |
| 3 | S | 3 | 5 | 15 | 5 | 3 | 1 |
| 4 | S | 3 | 6 | 18 | 6 | 3 | 1 |

Again, the main benefit of using the auxiliary views is that we can avoid computing the product $R \times S$ (or in general, joins) by simply summing up views. In this example, the view values of the $(k+1)$-th row can be computed by just three pairwise additions of values from the $k$-th row. □

This is the simplest possible example query for which the viewlet transform includes a second-order delta query, omitting any complex query features (e.g., predicates, self-joins, nested queries). Viewlet transforms can handle general update workloads including deletions and updates, as well as queries with multi-row results.

For a large fragment of SQL, *higher-order IVM avoids join processing in any form*, reducing all the view refreshment work to summation. Indeed, joins are only needed in the presence of inequality joins and nested aggregates in view definitions. The viewlet transform performs repeated (recursive) delta rewrites. Nested aggregates aside, each $k$-th order delta is structurally simpler than the $(k-1)$-th order delta query. The viewlet transform terminates, as for some $n$, the $n$-th order delta is guaranteed to be constant, only depending on the update but not on the database. In the above example, the second-order delta is constant, not using any database relation.

Our higher-order IVM framework, DBToaster, realizes as-incremental-as-possible query evaluation over SQL with a query language extending the bag relational algebra, query compilation and a variety of novel materialization and optimization strategies. DBToaster bears the promise of providing materialized views of complex long-running SQL queries, without window semantics or other restrictions, at very high view refresh rates. The data may change rapidly, and still part of it may be long-lived. A DDMS can use this functionality as the basis for richer query constructs than those supported by stream engines. DBToaster takes as input SQL view queries, and automatically incrementalizes them into procedural C++ trigger code where all work reduces to fine-grained, low-cost updates of materialized views.

**Example 2** Consider the query

```
Q = select sum(LI.PRICE * O.XCH)
from Orders O, LineItem LI where O.ORDK = LI.ORDK;
```

on a TPC-H like schema of Orders and Lineitem in which lineitems have prices and orders have currency exchange rates. The query asks for total sales across all orders weighted by exchange rates. We materialize the views for query $Q$ as well as the first-order views Q_LI ($\Delta_{LI}Q$) and Q_O ($\Delta_O Q$). The second-order deltas are constant w.r.t. the database and have been inlined in the following insert trigger programs that our approach produces for query $Q$.

```
on insert into O values (ordk, custk, xch) do {
   Q += xch * Q_O[ordk];
   Q_LI[ordk] += xch;
}
on insert into LI values (ordk, partk, price) do {
   Q += price * Q_LI[ordk];
   Q_O[ordk] += price;
}
```

The query result is again scalar, but the auxiliary views are not, and our language generalizes them from SQL's multi-sets to maps that associate multiplicities with tuples. This is again a very simple example (more involved ones are presented throughout the paper), but it illustrates something notable: while classical incremental view maintenance has to evaluate the first-order deltas, which takes linear time (e.g., $(\Delta_O Q)[\text{ordk}]$ is `select sum(LI.PRICE) from Lineitem LI where LI.ORDK=ordk`), we get around this by performing IVM of the deltas. This way our triggers can be evaluated in *constant* time for single-tuple inserts in this example.

The delete triggers for $Q$ are the same as the insert triggers with `+=` replaced by `-=` everywhere. □

This example presents single-tuple update triggers. Viewlet transforms are not limited to this but support bulk updates. However, delta queries for single-tuple updates have considerable additional optimization potential, which our compiler leverages to create very efficient code that refreshes views whenever a new update arrives. We do not queue updates for bulk processing, and so maximize view availability and minimize view refresh latency, enabling ultra-low latency monitoring and algorithmic trading applications.

On paper, higher-order IVM clearly dominates classical IVM. If classical IVM is a good idea, then doing it recursively is an even better idea. The same efficiency improvement argument in favor of IVM of the base query also holds for IVM of the delta query. Considering that joins are expensive and this approach eliminates them, higher-order IVM has the potential for excellent query performance.

In practice, how well do our expectations of higher-order IVM translate into real performance gains? A priori, the costs associated with storing and managing additional auxiliary materialized views for higher-order delta queries might be more considerable than expected. This paper presents the lessons learned in an effort to realize higher-order IVM, and to understand its strengths and drawbacks. Our contributions are as follows:

1. We present the concept of higher-order IVM and describe the viewlet transform. This part of the paper generalizes and consolidates our earlier work [3, 19].

2. There are cases (inequality joins and certain nesting patterns) when a naive viewlet transform is too aggressive, and certain parts of queries are better re-evaluated than incrementally maintained. We develop heuristics and a cost-based optimization framework for trading off materialization and lazy evaluation for best performance.

3. We have built the DBToaster system which implements higher-order IVM. It combines an optimizing compiler that creates efficient update triggers based on the techniques discussed above, and a runtime system (currently single-core and main-memory based[1]) to keep views continuously fresh as updates stream in at high data rates.

4. We present the first set of extensive experimental results on higher-order IVM obtained using DBToaster. Our experiments indicate that frequently, particularly for queries that consist of many joins or nested aggregation subqueries, our compilation approach dominates the state of the art, often by multiple orders of magnitude. On a workload of automated trading and ETL queries, we show that current systems cannot sustain fresh views at the rates required in algorithmic trading and real-time analytics, while higher-order IVM takes a significant step to making these applications viable.

Most of our benchmark queries contain features such as nested subqueries that no commercial IVM implementation supports, while our approach handles them all.

## 2. RELATED WORK

### 2.1 A Brief Survey of IVM Techniques

Database view management is a well-studied area with over three decades of supporting literature. We focus on the aspects of view materialization most pertinent to DDMS.

**Incremental Maintenance Algorithms, and Formal Semantics.** Maintaining query answers has been considered under both the set [6, 7] and bag [8, 14] relational algebra. Generally, given a query on $N$ relations $Q(R_1, \ldots, R_N)$, classical IVM uses a first-order delta query $\Delta_{R_1} Q = Q(R_1 \cup \Delta R_1, R_2, \ldots R_N) - Q(R_1, \ldots, R_N)$ for each input relation $R_i$ in turn. The creation of delta queries has been studied for query languages with aggregation [25] and bag semantics [14], but we know of no work that investigates delta queries of nested and correlated subqueries. [17] has considered view maintenance in the nested relational algebra (NRA), however this has not been widely adopted in any commercial DBMS. Finally, [33] considered temporal views, and [22] outer joins and nulls, all for flat SPJAG queries without generalizing to subqueries, the full compositionality of SQL, or the range of standard aggregates.

**Materialization and Query Optimization Strategies.** Selecting queries to materialize and reuse during processing has spanned fine-grained approaches from subqueries [27] and partial materialization [20, 28], to coarse-grained methods as part of multiquery optimization and common subexpressions [16, 35]. Picking views from a workload of queries typically uses the AND-OR graph representations from multiquery optimization [16, 27], or adopts signature and subsumption methods for common subexpressions [35]. [27] picks additional views to materialize amongst subqueries of the view definition, but only performs first-order maintenance and does not consider the full framework (binding patterns, etc) required with higher-order deltas. Furthermore, the optimal set of views is chosen based on the maintenance costs alone, from a search space that can be doubly exponential in the number of query relations.

Physical DB designers [2, 36] often use the query optimizer as a subcomponent to manage the search space of equivalent views, reusing its rewriting and pruning mechanisms. For partial materialization methods, ViewCache [28] and DynaMat [20] use materialized view fragments, the former materializing join results by storing pointers back to input tuples, the latter subject to a caching policy based on refresh time and cache space overhead constraints.

**Evaluation Strategies.** To realize efficient maintenance with first-order delta queries, [9, 34] studied eager and lazy evaluation to balance query and update workloads, and as background asynchronous processes [29], to achieve a variety of view freshness models and constraints [10]. Evaluating maintenance queries has also been studied extensively in Datalog with semi-naive evaluation (which also uses first-order deltas) and DRed (delete-rederive) [15]. Finally, [13] argues for view maintenance in stream processing, which reinforces our position of IVM as a general-purpose change propagation mechanism for collections, on top of which window and pattern constructs can be defined.

### 2.2 Update Processing Mechanisms

**Triggers and Active Databases.** Triggers, active database and event-condition-action (ECA) mechanisms [4] provide general purpose reactive behavior in a DBMS. The literature considers recursive and cascading trigger firings, and restrictions to ensure restricted propagation. Trigger-based approaches require developers to manually convert queries to delta form, a painful and error-prone process especially in the higher-order setting. Without manual incrementalization, triggers suffer from poor performance and cannot be optimized by a DBMS when written in C or Java.

**Data Stream Processing.** Data stream processing [1, 24] and streaming algorithms combine two aspects of handling updates: i) shared, incremental processing (e.g. sliding windows, paired vs paned windows), ii) sublinear algorithms (i.e. polylogarithmic space bounds). The latter are approximate processing techniques that are difficult to program and compose, and have had limited adoption in commercial DBMS. Advanced processing techniques in the streaming community also focus almost entirely on approximate techniques when processing cannot keep up with stream rates (e.g. load shedding, prioritization [30]), on shared processing (e.g. on-the-fly aggregation [21]), or specialized algorithms and data structures [11]. Our approach to streaming is about generalizing incremental processing to (non-windowed) SQL semantics (including nested subqueries and aggregates). Of course, windows can be expressed in this semantics if desired. Similar principles are discussed in [13].

**Automatic Differentiation and Incrementalization, and Applications.** Beyond the database literature, the programming language literature has studied automatic incrementalization [23], and automatic differentiation. Automatic incrementalization is by no means a solved challenge, especially when considering general recursion and unbounded iteration. Automatic differentiation considers deltas of functions applied over scalars rather than sets or collections, and lately in higher-order fashion [26]. Bridging these two areas of research would be fruitful for DDMS to support UDFs and general computation on scalars and collections.

## 3. QUERIES AND DELTAS

In this section, we fix a formalism for queries that makes it easy to talk cleanly and concisely about delta processing, and we describe the construction of delta queries.

### 3.1 Data Model and Query Language

Our data model generalizes multiset relations (as in SQL) to tuple multiplicities that are rational numbers. This for one allows us to treat databases and updates uniformly (for instance, a delete is a relation with negative multiplicities, and applying it to a database means unioning/adding it to the database). It also allows us to use multiplicities to represent aggregate query results (which do not need to be integers). As a consequence, when performing delta-processing on aggregate queries, growing an aggregate means adding to the aggregate value rather than to delete the tuple with the old aggregate value and insert a tuple with the new aggregate value. Maintaining aggregates in the multiplicities allows for simpler and cleaner bookkeeping. It is a cosmetic change that does not keep us from supporting SQL queries.

---
[1] This is not an intrinsic limitation of our method, in fact our trigger programs are particularly nicely parallelizable [19].

| $R$ | $\langle A, B\rangle$ | # |
|---|---|---|
| | $\langle a, b_1\rangle \mapsto$ | 2 |
| | $\langle a, b_2\rangle \mapsto$ | −3 |

| $S_1$ | $\langle B, C\rangle$ | # |
|---|---|---|
| | $\langle b_1, c_1\rangle \mapsto$ | 2 |
| | $\langle b_1, c_2\rangle \mapsto$ | −10 |

| $S_2$ | $\langle B, C\rangle$ | # |
|---|---|---|
| | $\langle b_1, c_1\rangle \mapsto$ | 3 |
| | $\langle b_1, c_2\rangle \mapsto$ | 3 |
| | $\langle b_2, c_1\rangle \mapsto$ | −11 |

| $S_1 + S_2$ | $\langle B, C\rangle$ | # |
|---|---|---|
| | $\langle b_1, c_1\rangle \mapsto$ | 5 |
| | $\langle b_1, c_2\rangle \mapsto$ | −7 |
| | $\langle b_2, c_1\rangle \mapsto$ | −11 |

| $R \bowtie (S_1 + S_2)$ | $\langle A, B, C\rangle$ | # |
|---|---|---|
| | $\langle a, b_1, c_1\rangle \mapsto$ | 10 |
| | $\langle a, b_1, c_2\rangle \mapsto$ | −14 |
| | $\langle a, b_2, c_1\rangle \mapsto$ | 33 |

| $\text{Sum}_{AC;1/2}(R \bowtie (S_1 + S_2))$ | $\langle A, C\rangle$ | # |
|---|---|---|
| | $\langle a, c_1\rangle \mapsto$ | 21.5 |
| | $\langle a, c_2\rangle \mapsto$ | −7 |

**Figure 1: Examples of union, join, and aggregation of generalized multiset relations with rational number tuple multiplicities.**

Formally, a thus *generalized multiset relation* (GMR) is a function from tuples to rational numbers, with finite support (i.e., only a finite number of tuples have nonzero multiplicity). The union, join, selection, and grouping-sum-aggregate operations are defined in the way that naturally generalizes the same operations on multiset relations:

$$R + S \;:\; \vec{t} \mapsto R(\vec{t}) + S(\vec{t})$$

$$R \bowtie S \;:\; \vec{t} \mapsto \begin{cases} R(\pi_{\text{sch}(R)}\vec{t}) * S(\pi_{\text{sch}(S)}\vec{t}) \\ \qquad \ldots \; \vec{t} = \pi_{\text{sch}(R) \cup \text{sch}(S)}(\vec{t}) \\ 0 \quad \ldots \text{ otherwise} \end{cases}$$

$$\sigma_\theta R \;:\; \vec{t} \mapsto \begin{cases} R(\vec{t}) & \ldots \; \theta(\vec{t}) \text{ is true} \\ 0 & \ldots \text{ otherwise} \end{cases}$$

$$\text{Sum}_{\vec{A};f} R \;:\; \vec{a} \mapsto \sum_{\pi_{\vec{A}}(\vec{t}) = \vec{a}} R(\vec{t}) * f(\vec{t})$$

Here, $\pi$ is the projection of records, rather than relations, removing fields of $\vec{t}$ whose labels are not among the column names $\vec{A}$, and $\text{sch}(R)$ denotes the list of column names of GMR $R$. An aggregation $\text{Sum}_{\vec{A};f} R$ almost works like the SQL query `select` $\vec{A}$`, sum(f) from R group by` $\vec{A}$, with the difference that SQL puts the aggregate value into a new column, while $\text{Sum}_{\vec{A};f} R$ puts it into the multiplicity of the group-by tuple. Aggregations can also serve as multiplicity-preserving projections, and a query $\text{Sum}_{\vec{A};1}(R)$ corresponds equivalently to SQL queries `select` $\vec{A}$ `from R` and `select` $\vec{A}$`, sum(1) from R group by` $\vec{A}$. Examples of join, union, and aggregation are shown in Figure 1.

Our query language includes relational atoms $R$, constant singleton relations, natural join, union (denoted $+$), selection, grouping sum-aggregates, and column renaming $\rho$:

$$Q ::= R \mid \{\vec{A} : \vec{a} \mapsto c\} \mid Q \bowtie Q \mid Q + Q \mid \sigma_\phi Q \mid \text{Sum}_{\vec{A};f} Q \mid \rho_{\vec{A}} Q$$

where the $c$ are rational numbers, $f$ are terms, and $\phi$ are conditions over terms. Terms are define using arithmetics over rational constants and column names. Additionally we can use non-grouping aggregates as terms (the value is the multiplicity), specifically in selection conditions. That way we can express queries with nested aggregates. Nested aggregates may be correlated with the outside in the way we are used to from SQL. For example, we write $\sigma_{C < \text{Sum}_{A;B} R} S$ for the SQL query

```
select * from S
where S.C < (select sum(B) from R where R.A=S.A)
```

We can perform deletions writing $R - S$, and this is no fundamentally new operation since we can define $R - S := R + (S \bowtie \{\langle\rangle \mapsto -1\})$. Here $\{\langle\rangle \mapsto -1\}$ is a nullary case of singleton GMR construction $\{\vec{A} : \vec{a} \mapsto c\}$.

There is no explicit syntax for universal quantification/relational difference or aggregates other than Sum, but all these features can be expressed using (nested) sum-aggregate queries (a popular homework exercise in database courses). Special handling of these features in delta processing and query optimization could yield performance better than what we report in our experiments. However, granting these definable features specialized treatment is beyond the scope of this paper. As a consequence, our implementation provides native support for only the fragment presented above and the experiment use only techniques described in the paper.

We will use relational algebra and SQL syntax interchangeably as we are used to from bag relational algebra and SQL.

## 3.2 Computing the Delta of a Query

We next show how to construct delta queries. The reader familiar with incremental view maintenance may skip this section, but note that the algebra just fixed has the nice property of being *closed under taking deltas*. For each query expression $Q$, there is an expression $\Delta Q$ of the same algebra that expresses how the result of $Q$ changes as the database $D$ is changed by update workload $\Delta D$,

$$\Delta Q(D, \Delta D) := Q(D + \Delta D) - Q(D).$$

Thanks to the strong compositionality of the language, we only have to give delta rules for the individual operators. These rules are given and studied in detail in [19]. In short,

$$\begin{aligned}
\Delta(Q_1 + Q_2) &:= (\Delta Q_1) + (\Delta Q_2), \\
\Delta(\text{Sum}_{\vec{A};f} Q) &:= \text{Sum}_{\vec{A};f}(\Delta Q), \\
\Delta(Q_1 \bowtie Q_2) &:= ((\Delta Q_1) \bowtie Q_2) + (Q_1 \bowtie (\Delta Q_2)) \\
&\quad + ((\Delta Q_1) \bowtie (\Delta Q_2)), \\
\Delta(\sigma_\theta Q) &:= \sigma_\theta(\Delta Q).
\end{aligned}$$

$\Delta R$ is the update to $R$. In the case that the update does not change $R$ (but other relation(s)), $\Delta R$ is empty.

Here we assume that $f$ and $\theta$ do not contain nested aggregates. Achieving the full generality is not a technical problem, but requires notation beyond the scope of this paper; we refer to [19] for the general case. We will see in Section 5 that, in practice, we will not need deltas for conditions with nested aggregates, as we will decide to re-evaluate rather than materialize and incrementally maintain such conditions.

**Example 3** Given schema $R(AB), S(CD)$, and query

```
select sum(A * D) from R, S where B = C
```

or, in the algebra, $\text{Sum}_{\langle\rangle;A*D}(\sigma_{B=C}(R \bowtie S))$. Modulo names, this is the query of Example 2. The delta for this query as

971

we apply a change $\Delta R$ to relation $R$ but leave $S$ unchanged ($\Delta S$ is empty) is

$$\Delta\text{Sum}_{\langle\rangle;A*D}(\sigma_{B=C}(R \bowtie S)) =$$
$$\text{Sum}_{\langle\rangle;A*D}\Delta(\sigma_{B=C}(R \bowtie S)) =$$
$$\text{Sum}_{\langle\rangle;A*D}(\sigma_{B=C}\Delta(R \bowtie S))$$

and by the delta rule for $\bowtie$,

$$\Delta(R \bowtie S) = (\Delta R) \bowtie S + R \bowtie (\Delta S) + (\Delta R) \bowtie (\Delta S).$$

Thus the delta query is $\text{Sum}_{\langle\rangle;A*D}(\sigma_{B=C}((\Delta R) \bowtie S))$. □

The delta rules work for bulk updates. The special case of single-tuple updates is interesting since it allows us to simplify delta queries further and to generate particularly efficient view refresh code.

**Example 4** We continue the Example 3, but now assume that $\Delta R$ is an insertion of a single tuple $\langle A:x, B:y \rangle$. The delta query $\text{Sum}_{\langle\rangle;A*D}(\sigma_{B=C}(\{\langle A:x,B:y\rangle\} \bowtie S))$ can be simplified to $\text{Sum}_{\langle\rangle;x*D}(\sigma_{y=C}S)$. □

## 3.3 Binding Patterns

Query expressions have binding patterns: There are input variables or parameters without which we cannot evaluate these expressions, and there are output variables, the columns of the schema of the query result. Each expression $Q$ has input variables or parameters $\vec{x_{in}}$ and a set of output variables $\vec{x_{out}}$, which form the schema of the query result. We denote such an expression as $Q[\vec{x_{in}}][\vec{x_{out}}]$. The input variables are those that are not *range-restricted* in a calculus formulation, or equivalently have to be understood as *parameters* in an SQL query because their values cannot be computed from the database: They have to be provided so that the query can be evaluated.

The most interesting case of input variables occurs in a correlated nested subquery, viewed in isolation. In such a subquery, the correlation variable from the outside is such an input variable. The subquery can only be computed if a value for the input variable is given.

**Example 5** We illustrate this by an example. Assume relation $R$ has columns $A, B$ and relation $S$ has columns $C, D$. The SQL query

```
select * from R
where B < (select sum(D) from S where A > C)
```

is equivalent to $\text{Sum}_{*;1}(\sigma_{B<\text{Sum}_{\langle\rangle;D}(\sigma_{A>C}(S))}R)$ in the algebra. Here, all columns of the schema of $R$ are output variables. In the subexpression $\text{Sum}_{\langle\rangle;D}(\sigma_{A>C}(S))$, $A$ is an input variable; there are no output variables since the aggregate is non-grouping. □

Also note that taking a delta *adds input variables* parameterizing the query with the update. In Example 4 for instance, the delta query has input variables $x$ and $y$ to pass in the update. Delta queries for bulk updates have relation-valued parameters.

## 4. THE VIEWLET TRANSFORM

We are now ready for the viewlet transform. If we restrict the query language to exclude aggregates nested into conditions[2] (for which the delta query was complicated), the query language fragment has the following nice property.

---
[2]We will eliminate this restriction on the technique in the next section.

$\Delta Q$ is structurally strictly simpler than $Q$ when query complexity is measured as follows. For union(+)-free queries, the *degree* $\deg(Q)$ of query $Q$ is the number of relations joined together. We can use distributivity to push unions above joins and so give a degree to queries with unions: the maximum degree of the union-free subqueries. Queries are strongly analogous to polynomials, and the degree of queries is defined precisely as it is defined for polynomials (where the relation atoms of the query correspond to the variables of the polynomial).

**Theorem 1 ([19])** *If* $\deg(Q) > 0$, *then*

$$\deg(\Delta Q) = \deg(Q) - 1.$$

The viewlet transform makes use of the simple fact that a delta query is a query too. Thus it can be incrementally maintained, making use of a delta query to the delta query, which again can be materialized and incrementally maintained, and so on, recursively. By the above theorem, this recursive query transformation terminates in the $\deg(Q)$-th recursion level, when the rewritten query is a "constant" independent of the database, and dependent only on updates.

All queries, aggregate or not, map tuples to rational numbers (= define GMRs). Thus it is natural to think of the views as map data structures (dictionaries). In this section, we make no notational difference between queries and (materialized) views, but it will be clear from the context that we are using views when we increment them.

**Definition 1** The viewlet transform turns a query into a set of update triggers that together maintain the view of the query and a set of auxiliary views. Assume the query $Q$ has input variables (parameters) $\vec{x}_{in}$. For each relation $R$ used in the query, the viewlet transform creates a trigger

```
on update R values D_R do T_R.
```

where $D_R$ is the update – a generalized multiset relation – to the relation named $R$ and $T_R$ is the trigger body, a set of statements. The trigger bodies are most easily defined by their computation by a recursive imperative procedure $VT$ defined below (initially, the statement lists $T_R$ are empty):

procedure $\text{VT}(Q, \vec{x}_{in})$:
foreach relation name $R$ used in $Q$ do {
    $\text{T}_R$ += (`foreach` $\vec{x}_{in}$ `do` $Q[\vec{x}_{in}]$ += $\Delta_R Q[\vec{x}_{in} D_R]$)
    if $\deg(Q) > 0$ then {
        let $D$ be a new variable of type relation of schema $R$;
        $\text{VT}(\Delta_R Q, \vec{x}_{in} D)$
} }

Here, += on $T_R$ appends a statement to an imperative code block, and += on generalized multiset relations uses the + of Section 3.1. Exactly those queries occurring in triggers that have degree greater than zero are materialized. Of course, these are exactly those queries that are added to by trigger statements: those that are incrementally maintained. □

**Example 6** For example, if the schema contains two relations $R$ and $S$ and query $Q$ has degree 2, then $\text{VT}(Q, \langle\rangle)$ returns as $T_R$ the code block

```
Q += Δ_R Q[D_R];
foreach D_1 do Δ_R Q[D_1] += Δ_R Δ_R Q[D_1, D_R];
foreach D_2 do Δ_S Q[D_2] += Δ_R Δ_S Q[D_2, D_R]
```

The body for the update trigger of $S$, $T_S$, is analogous. Note that the order of the first two statements matters. For correctness, we read the old versions of views in a trigger. □

The viewlet transform bears a superficial analogy with the Haar wavelet transform, which also materializes a hierarchy of differences; however, these are not differences of differences but differences of recursively computed sums.

At runtime, each trigger statement loops over a relevant subset of the possible valuations of the parameters of the views used in the statement. For relation-typed parameters, this is a priori astronomically expensive. There are various ways of bounding the domains to make this feasible. Furthermore, parameters can frequently be optimized away. Nevertheless, single-tuple updates offer particular optimization potential, and we focus on these in this paper.

We will study single-tuple insertions, denoted $+R(\vec{t})$ for the insertion of tuple $\vec{t}$ into relation $R$, and single-tuple deletions $-R(\vec{t})$. Here we create insert and delete triggers in which the argument is the tuple, rather than a generalized multiset relation, and we avoid looping over relation-typed variables.

**Example 7** We return to the query $Q$ of Example 4, with single-tuple updates. The query has degree two. The second-order deltas $(\Delta_{\text{sgn}_R R(x,y)}\Delta_{\text{sgn}_S S(z,u)}Q)[x,y,z,u]$ have value $\text{sgn}_R\text{sgn}_S \text{Sum}_{\langle\rangle;x*u}(\sigma_{y=z}\{\langle\rangle\})$, which is equivalent to $\langle\rangle \mapsto \text{sgn}_R\text{sgn}_S$ if $y = z$ then $x * u$ else 0; here $\text{sgn}_R, \text{sgn}_S \in \{+,-\}$. Variables $x$ and $y$ are arguments of the trigger and are bound at runtime, but variables $z$ and $u$ need to be looped over. On the other hand, the right-hand side of the trigger is only non-zero in case that $y = z$. So we can substitute $z$ by $y$ everywhere and eliminate $z$. Using this simplification, the on-insert into $R$ trigger $+R(x,y)$ according to the viewlet transform is the program

```
Q += Δ_{+R(x,y)}Q[x, y];
foreach u do Δ_{+S(y,u)}Q[y, u] += {⟨⟩ ↦ x * u};
foreach u do Δ_{-S(y,u)}Q[y, u] -= {⟨⟩ ↦ x * u}
```

The remaining triggers are constructed analogously. The trigger contains an update rule for the (in this case, scalar) view $Q$ for the overall query result, which uses the auxiliary view $\Delta_{\pm R(x,y)}Q$ which is maintained in the update triggers for $S$, plus update rules for the auxiliary views $\Delta_{\pm S(z,u)}Q$ that are used to update $Q$ in the insertion and deletion triggers on updates to $S$.

The reason why we did not show deltas $\Delta_{\pm R(\ldots)}\Delta_{\pm R(\ldots)}Q$ or $\Delta_{\pm S(\ldots)}\Delta_{\pm S(\ldots)}Q$ is that these are guaranteed to be 0, as the query does not have a self-join.

A further optimization, exploiting distributivity and presented in the next section, eliminates the loops on $u$ and leads us to the triggers of Example 2. □

We observe that the structure of the work that needs to be done is extremely regular and (conceptually) simple. Moreover, there are no classical large-granularity operators left, so it does not make sense to give this workload to a classical query optimizer. There are for-loops over many variables, which have the potential to be very expensive. But the work is also perfectly data-parallel, and there are no data dependencies comparable to those present in joins. All this provides justification for making heavy use of compilation.

Note that there are many optimizations not exploited in the presentation of viewlet transforms. Thus, we will refer to the viewlet transform as presented in this section as *naive recursive IVM* in the experiments section. We will present improvements and optimizations next.

## 5. OPTIMIZING VIEWLETS

In this section, we present optimizations of the viewlet transform as well as heuristics and a cost model that allow us to avoid materializing views with high maintenance cost.

**Query Decomposition**

$\mathcal{M}(\text{Sum}_{\vec{A}\vec{B}:f_1*f_2}(Q_1 \bowtie Q_2)) \Rightarrow$

$$\mathcal{M}(\text{Sum}_{\vec{A}:f_1}(Q_1)) \bowtie \mathcal{M}(\text{Sum}_{\vec{B}:f_2}(Q_2)) \quad (1)$$

$Q_1$ and $Q_2$ have no common columns
$\vec{A}$ and $\vec{B}$ are the group-by terms of each

**Factorization and Polynomial Expansion**

$\mathcal{M}(Q_L \bowtie (Q_1 + Q_2 + \ldots) \bowtie Q_R) \Leftrightarrow$

$$\mathcal{M}(Q_L \bowtie Q_1 \bowtie Q_R) + \mathcal{M}(Q_L \bowtie Q_2 \bowtie Q_R) + \ldots \quad (2)$$

**Input Variables**

$\mathcal{M}(\text{Sum}_{\vec{A};f(\vec{B}\vec{C})}(\sigma_{\theta(\vec{B}\vec{C})}(Q))) \Rightarrow$

$$\text{Sum}_{\vec{A};f(\vec{B}\vec{C})}(\sigma_{\theta(\vec{B}\vec{C})}(\mathcal{M}(\text{Sum}_{\vec{A}\vec{B};1}(Q)))) \quad (3)$$

$f, \theta$ are functions over terms
$\vec{A}$ is the group-by variables of the aggregate over $Q$
$\vec{B}$ is the output variables of $Q$ used by $f, \theta$
$\vec{C}$ is the input variables that do not appear in $Q$

**Nested Aggregates and Decorrelation**

$\mathcal{M}(\text{Sum}_{\vec{A};f}(\sigma_{\theta(Q_N,\vec{B})}(Q_O))) \Rightarrow$

$$\text{Sum}_{\vec{A};1}(\sigma_{\theta(\mathcal{M}(Q_N),\vec{B})}(\mathcal{M}(\text{Sum}_{\vec{A}\vec{B};f}(Q_O)))) \quad (4)$$

$Q_N$ is a non-grouping aggregate term
$f, \theta$ are functions over terms
$\vec{A}$ is the group-by variables of the aggregate over $Q_O$
$\vec{B}$ is the output variables of $Q_O$ used by $f, \theta$ and $Q_N$

**Figure 2: Rewrite rules for partial materialization. Bidirectional arrows indicate rules that are applied heuristically or using the cost model from Section 5.1.**

### 5.1 Materialization Decisions

For any query $Q$, the naive viewlet transform produces a single materialized view $M_Q$. However, it is frequently more efficient to materialize $Q$ piecewise, as a collection of incrementally maintained materialized views $\vec{M}_Q$ and an equivalent query $Q'$ that is evaluated over these materialized views. We refer to the rewritten query and its piecewise maps as a *materialization decision* for Q, denoted $\langle Q', \vec{M}_Q \rangle$.

DBToaster selects a materialization decision for a query $Q$ iteratively, by starting with the naive materialization decision $\langle (M_{Q,1}), (M_{Q,1} := Q) \rangle$ and applying several rewrite rules up to a fixed point. These rewrite rules are presented in Figure 2, and are discussed individually below. Figure 3 shows the applicability of these rules to the experimental workload discussed in Section 6 and Appendix A.

For clarity, we will use a materialization operator $\mathcal{M}$ to show materialization decisions juxtaposed with their corresponding queries. For example, one possible materialization decision for the query $Q := Q_1 \bowtie Q_2$ is:

$$\mathcal{M}(Q_1) \bowtie \mathcal{M}(Q_2) \equiv \langle (M_{Q,1} \bowtie M_{Q,2}), \{M_{Q,i} := Q_i\} \rangle.$$

We will first discuss the use of these rules in a heuristic optimizer, which when applied as aggressively as possible produces near-optimal trigger programs for maintaining the vast majority of queries. We also briefly discuss a cost-based optimization strategy to further improve performance.

**Duplicate View Elimination.** As the simplest optimization, we observe that the naive viewlet transform produces many duplicate views. This is primarily because the delta operation typically commutes with itself; $\Delta_R\Delta_S Q = \Delta_S\Delta_R Q$ for any $Q$ that does not contain nested aggregates over $R$ or $S$. Even simple structural equivalence effectively identifies this sort of view duplication. View de-duplication substantially reduces the number of views created.

**Query Decomposition.** DBToaster makes extensive use of the generalized distributive law[5] (which plays an important role for probabilistic inference with graphical models) to decompose the materialization of expressions with disconnected join graphs. This rule is presented in Figure 2.1.

If the join graph of $Q$ includes multiple disconnected components $\vec{Q}_i$, $Q_2$, … (i.e., $Q := Q_1 \times Q_2 \times \ldots$), it is better to materialize each component independently. The cost of selecting from, or iterating over $Q$ is identical for both materialization strategies. Furthermore, maintaining each individual $Q_i$ is less computationally expensive; the decomposed materialization stores (and maintains) only $\sum_i |Q_i|$ values, while the combined materialization handles $\prod_i |Q_i|$ values.

This optimization plays a major role in the efficiency of DBToaster, and in the justification of compilation. Taking a delta of a query with respect to a single-tuple update replaces a relation in the query by a singleton constant tuple, effectively eliminating one hyperedge from the join graph and creating new disconnected components that can be further decomposed. Query decomposition is also critical for ensuring that the number of maps created for any acyclic query is polynomial.

**Polynomial Expansion and Factorization.** As described above, query decomposition operates exclusively over conjunctive queries. In order to decompose across unions, we observe that the union operation can be pushed through aggregate sums (i.e, $\text{Sum}(Q_1 + Q_2) = \text{Sum}(Q_1) + \text{Sum}(Q_2)$), and distributed over joins.

Any query expression can be expanded into a flattened *polynomial* representation, which consists of a union of purely conjunctive queries. Query decomposition is then applied to each conjunctive query individually. This rewriting rule is presented in Figure 2.2.

Note that this rewrite rule can also be applied in reverse. A polynomial expression can be *factorized* into a smaller representation by identifying common subexpressions ($Q_L$ and $Q_R$ in the rewrite rule) and pulling them out of the union. The cost-based optimizer makes extensive use of factorization to fully explore the space of possible materialization decisions. The heuristic optimizer does not attempt to factorize while making a materialization decision. However, after a materialization decision $\langle M_Q, \{\ldots\}\rangle$ is finalized, the materialized query $M_Q$ itself is simplified by factorization.

**Input Variables.** Queries with input variables have infinite domains and cannot be fully materialized. By default, DBToaster's heuristics ensures that input variables are avoided, in the query and all its subexpressions.

Input variables are originally introduced into an expression by nested aggregates, and as part of deltas. They appear exclusively in selection and aggregation operators.

The rewrite rule shown in Figure 2.3 ensures that materialized expressions do not have input variables by pulling operators with input variables out of the materialized expression. If an operator can be partitioned into components with output variables only and components with input variables, only the latter are pulled out of the expression.

In addition to this heuristic approach, the cost-based optimizer explores alternative materialization decisions where some (or all) of the input variables in an expression are kept in the materialized expression. With respect to Figure 2.3, these input variables are treated as elements of $\vec{A}$ instead of $\vec{C}$. At runtime, only a finite portion of the domain of these input variables is maintained in the materialized view.

A materialized view with input variables acts, in effect, as a cache for the query results. Unlike a traditional cache however, the contents of this map are not invalidated when the underlying data changes, but instead maintained incrementally. These sorts of *view caches* are analogous to partially materialized views [20, 28].

View caches are only beneficial when the size of the active domain of an input variable is small, and so the heuristic optimizer does not attempt to create them[3].

**Deltas of Nested Aggregates.** Thus far, we have ignored queries containing nested subqueries. When the delta of the nested subquery is nonzero, the delta of the entire query is not simpler than the original. The full delta rule for nested presented in [19] effectively computes the nested aggregate twice: once to obtain the original value, and once to obtain the value after incorporating the delta.

**Example 8** Consider the following query, with relations $R$ and $S$ with columns $A$ and $B$ respectively:

$$Q := \text{Sum}_{\langle\rangle;1}(\sigma_{\text{Sum}_{\langle\rangle;1}(S)=A}(R))$$

By the delta rule for nested aggregates,

$$\Delta_{+S(B')} := \text{Sum}_{\langle\rangle;1}(\sigma_{\text{Sum}_{\langle\rangle;1}(S)+1=A}(R)) - \text{Sum}_{\langle\rangle;1}(\sigma_{\text{Sum}_{\langle\rangle;1}(S)=A}(R))$$

Because the original nested query appears in the delta expression, the naive viewlet transform will not terminate here. To address this, the delta query is is decorrelated into separate materialized expressions for the nested subquery and the outer query. The rewrite rule of Figure 2.4 is applied twice (once to each instance). Each materialized expression now has a lower degree than the original query.

Although this rule is necessary for termination, it introduces a computation cost when evaluating the delta query. Note however, that this rule is only required (and thus, only used) when the delta of the nested subquery is nonzero.

**Example 9** Continuing Example 8, the materialization decision for $\Delta_{+S(B')}Q$ uses two materialized views: $M_{Q,1} := \text{Sum}_{\langle\rangle;1}(S)$ and $M_{Q,2} := R$, and uses each twice. However, $\Delta_{+R(A')}Q$ naturally has a lower degree than $Q$, and is thus materialized in its entirety.

Additionally, we observe that for some queries, it is more efficient to entirely recompute the query $Q$ on certain updates. Consider the general form of a nested aggregate:

$$Q := \text{Sum}_{\vec{A};f_1}(\sigma(\text{Sum}_{\vec{A}\vec{B};f_2}(Q_O)))$$

The delta of $Q$ evaluates two nearly identical expressions. Naively, computing the delta costs twice as much as the original query and so re-evaluation is more efficient. However, if the delta's trigger arguments bind one or more variables in $\vec{B}$, then the delta query only aggregates over a subset of the tuples in $Q_O$ and will therefore be faster. Based on this analysis, the heuristic optimizer decides whether to re-evaluate or incrementally maintain any given delta query.

**Cost Model.** While DBToaster's full cost-based optimizer is beyond the scope of this paper, we now briefly discuss its cost model. The dominant processing overheads are: (1) Updating maps with query results, and (2) Sum aggregates, which compound tuples into aggregate values.

The cost of a materialization decision $\langle Q', \vec{M}_Q\rangle$ includes both an evaluation component ($\text{cost}_e$) and a maintenance component ($\text{cost}_m$) for both the view materialized for the original query and all of the higher-order views.

---

[3]View caching is required for *any* materialized expression without finite support. This includes some forms of nested aggregates without input variables. A precise characterization of these expressions is beyond the scope of this paper.

| | Query | Rule (1) | Rule (2) | Rule (3) (S)ubquery (C)ache | Rule (4) (R)e-eval (I)ncremental |
|---|---|---|---|---|---|
| Finance | AXF | - | ✓ | S | - |
| | BSP | - | ✓ | - | - |
| | BSV | - | ✓ | - | - |
| | MST | - | ✓ | S | R,I |
| | PSP | - | ✓ | S | R,I |
| | VWAP | - | ✓ | C | R |
| TPCH | Q3 | ✓ | ✓ | - | - |
| | Q11 | - | - | - | - |
| | Q17 | ✓ | ✓ | S | I |
| | Q18 | ✓ | ✓ | S | I |
| | Q22 | ✓ | ✓ | S | R,I |
| | SSB4 | ✓ | - | - | - |

**Figure 3: Rewrite rules applied to our workload.**

DBToaster uses standard cardinality estimation [12, 32] for the number of distinct tuples when projecting the result of query $Q$ down to columns $\vec{A}$. We refer to this as the size of the domain of $\vec{A}$ in $Q$ ($|\text{dom}_{\vec{A}}(Q)|$). If $\vec{A}$ is the full set of output variables of $Q$, then we refer to this as the complete domain of $Q$ ($|\text{dom}(Q)|$).

Let $\mathbb{Q}$ be the set of all subexpressions of $Q$. The cost of query $Q$ is the sum of the sizes of the complete domains of the outer query $Q$ and all queries nested immediately inside aggregate sums $Q_i$ in $\mathbb{Q}$:

$$\text{cost}_e(Q) = |\text{dom}(Q)| + \sum_{\{Q_i | \text{Sum}(Q_i) \in \mathbb{Q}\}} |\text{dom}(Q_i)|$$

The maintenance cost of $Q$ is based on the cost of maintaining all the $M_{Q,i} \in \vec{M}_Q$. Every $M_{Q,i}$ must be updated for every change to a relation $R_j$ that appears in it. If the delta query of $M_{Q,i}$ is materialized using materialization decision $\langle Q'_{i,j}, \vec{M}_{Q_{i,j}}\rangle$, and the rate of insertions into $R_j$ is $\text{rate}_{R_j}$, then the cost of maintaining $M_{Q,i}$ is

$$\text{cost}_m(M_{Q,i}) = \sum_{R_j} \text{rate}_{R_j} \cdot \text{cost}_e(Q'_{i,j}) + \sum_{M \in \vec{M}_{Q_{i,j}}} \text{cost}_m(M)$$

This definition recurs on the cost of maintaining the maps required to evaluate $Q_{i,j}$. The maintenance cost of a map that is already being materialized by another query is zero.

The full cost of processing query $Q$ is now

$$\text{cost}(Q) = (\text{rate}_{\text{refresh}} \cdot \text{cost}_e(Q')) + (\sum_i \text{cost}_m(M_{Q,i}))$$

where the refresh rate of $Q$ depends on how frequently a fresh view must be made available. For the typical usage scenario a refresh occurs on every update, and $\text{rate}_{\text{refresh}} = \sum_j \text{rate}_{R_j}$.

### 5.2 Optimized Viewlet Transform Example

Figure 4 shows the trigger program compiled from query Q18 from our test workload (see Appendix A).

For simplicity, we use the condensed schema $C(CK)$, $O(CK, OK)$, and $LI(OK, QTY)$. The query $Q[][CK]$ is:

$$\text{Sum}_{CK;QTY}($$
$$\sigma_{100<\text{Sum}_{\langle\rangle;QTY'}(\sigma_{OK=OK'}(\rho_{OK',QTY'}LI))}(C \bowtie O \bowtie LI))$$

Due to space limitations we only show the derivation of insertions into Orders $O$ and Lineitem $LI$. Insertions into Customer $C$ are a simple extension, while deletions are duals of insertions and are omitted entirely.

**Insertions into Orders.** The first-order delta of $Q$ for insertion of a single tuple $\langle CK : ck, OK : ok \rangle$ is

$$\Delta_{+O\langle ck,ok\rangle}Q := \text{Sum}_{CK;QTY}(\sigma_{OK=ok \wedge CK=ck \wedge Q_{ns}}(C \bowtie LI))$$
$$\text{where } Q_{ns} = (100 < \text{Sum}_{\langle\rangle;QTY'}(\sigma_{ok=OK'}\rho_{OK',QTY'}LI))$$

```
on insert into C values (ck) do {
01   Q[][ck] += Q_C[][ck]
02   foreach OK do Q_{LI}[][ck,OK] += Q_{LI,C}[][ck,OK]
03   Q_{O1}[][ck] += 1
}
on insert into O values (ck,ok) do {
04   Q[][ck] += Q_{O1}[][ck] ⋈ Q_{O2}[][ok] ⋈ σ_{100<Q_{O2}[][ok]}(1)
05   Q_{LI}[][ck,ok] += Q_{O1}[][ck]
06   Q_{LI,C}[][ck,ok] += 1
07   Q_C[][ck] += Q_{O2}[][ok] ⋈ σ_{100<Q_{O2}[][ok]}(1)
}
on insert into LI values (ok,qty) do {
08   foreach CK do
       Q[][CK] += Q_{LI}[][CK,ok] ⋈ (
         ((Q_{O2}[][ok] + {⟨⟩ ↦ qty}) ⋈ σ_{100<qty+Q_{O2}[][ok]}(1))
         - (Q_{O2}[][ok] ⋈ σ_{100<Q_{O2}[][ok]}(1)))
09   foreach CK do
       Q_C[][CK] += Q_{LI,C}[][CK,ok] ⋈ (
         ((Q_{O2}[][ok] + {⟨⟩ ↦ qty}) ⋈ σ_{100<qty+Q_{O2}[][ok]}(1))
         - (Q_{O2}[][ok] ⋈ σ_{100<Q_{O2}[][ok]}(1)))
10   Q_{O2}[][ok] += qty
}
```

**Figure 4: DBToaster insert trigger program for Q18.**

By rewrite rule 1, this delta expression can be decomposed into two separate maps, since $C$ and $LI$ share no common columns. Furthermore, the nested subexpression does not contain relation $O$, so we do not apply rewrite rule 4 here. The delta expression can be materialized as follows:

$$\mathcal{M}(\text{Sum}_{\langle CK\rangle;1}(\sigma_{CK=ck}C)) \bowtie \mathcal{M}(\text{Sum}_{\langle\rangle;QTY}($$
$$\sigma_{OK=ok \wedge (100<\text{Sum}_{\langle\rangle;QTY'}(\sigma_{ok=OK'}(\rho_{OK',QTY'}LI)))}LI))$$

The second materialized map can be simplified further by rules 1 and 4. $OK$ is bound to trigger parameter $ok$, which breaks the join graph between the selection predicate and $LI$. Then, since the selection predicate is being applied to a singleton, we can safely materialize only the aggregate in the predicate. Applying these optimizations gives us the following materialization decision (with $1 := \{\langle\rangle \mapsto 1\}$):

$$\mathcal{M}(\text{Sum}_{\langle CK\rangle;1}(\sigma_{CK=ck}C)) \bowtie$$
$$\mathcal{M}(\text{Sum}_{\langle\rangle;QTY}(\sigma_{OK=ok}LI)) \bowtie$$
$$\text{Sum}_{\langle\rangle;1}(\sigma_{100<\mathcal{M}(\text{Sum}_{\langle\rangle;QTY}(\sigma_{ok=OK}LI))}(1))$$

Trigger statement 04 uses the following set of views (note that $Q_{O2}$ is used twice):

$$Q_{O1} := \text{Sum}_{CK;1}(C) \quad Q_{O2} := \text{Sum}_{OK;QTY}(LI)$$

$Q_{O1}[][CK]$ is maintained on insertions into $C$ with:

$$\Delta_{+C\langle ck\rangle}Q_{O1} := \{\langle CK : ck\rangle \mapsto 1\}$$

which corresponds to trigger statement 03. $Q_{O2}[][OK]$ is maintained similarly with trigger statement 10.

**Insertions into Lineitem.** The first-order delta of $Q$ for insertion of a single tuple $\langle OK : ok, QTY : qty\rangle$ is

$$\Delta_{+LI\langle ok,qty\rangle}Q := \text{Sum}_{CK;QTY}(\sigma_{OK=ok \wedge 100<qty+Q_{ns}}($$
$$C \bowtie O \bowtie (LI + \{\langle OK : ok, QTY : qty\rangle \mapsto 1\}))$$
$$- \sigma_{OK=ok \wedge 100<Q_{ns}}(C \bowtie O \bowtie LI))$$

When computing the above delta, we extend the delta rule of [19] for nested aggregates. The delta of a (decorrelated)

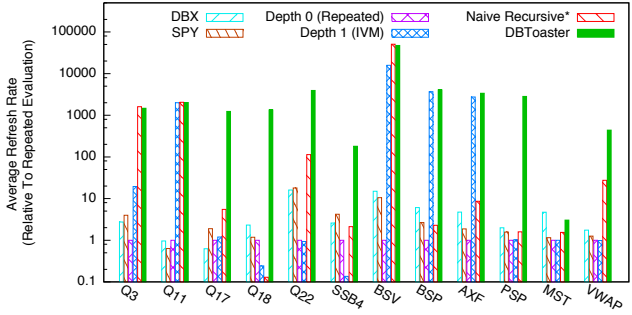

**Figure 5: DBToaster performance overview.** Note the logscale on the y-axis. (*) For VWAP, where DBToaster uses naive recursive compilation, we compare against a strategy that avoids input variables.

|  | Query | # Tables, join type. =: equi x: cross | Where-clause | Group-bys | # Subqueries and depth |
|---|---|---|---|---|---|
| Finance | AXF | 2, = | $\vee, <$ | yes | 0 / 0 |
|  | BSP | 2, = | $\wedge, <$ | yes | 0 / 0 |
|  | BSV | 2, = | None | no | 0 / 0 |
|  | MST | 2, x | $\wedge, <$ | yes | 2 / 1 |
|  | PSP | 2, x | $\wedge, <$ | no | 2 / 1 |
|  | VWAP | 1 | $<$ | no | 2 / 1 |
| TPCH | Q3 | 3, = | $\wedge, <$ | yes | 0 / 0 |
|  | Q11 | 2, = | None | yes | 0 / 0 |
|  | Q17 | 2, = | $<$ | no | 1 / 1 |
|  | Q18 | 3, = | $<$ | yes | 1 / 2 |
|  | Q22 | 1 | $=, <$ | yes | 2 / 1 |
|  | SSB4 | 7, = | $<$ | yes | 0 / 0 |

**Figure 6: Features of the algorithmic trading and online decision support workloads used for experiments.**

| Query | Depth 0 | DBX | DBX IVM | SPY | **DBToaster** |
|---|---|---|---|---|---|
| **TPCH3** | 15.00 | 41.12 | 4.01 | 59.69 | 22049.06 |
| **TPCH11** | 20.67 | 19.91 | 8.31 | 13.11 | 41842.62 |
| **TPCH17** | 17.15 | 10.72 | — | 32.58 | 21256.16 |
| **TPCH18** | 17.93 | 41.35 | — | 21.13 | 24056.02 |
| **TPCH22** | 2.22 | 35.65 | — | 39.82 | 8761.95 |
| **SSB4** | 16.49 | 42.67 | 6.30 | 69.19 | 2970.64 |
| **BSV** | 2.18 | 32.60 | 26.32 | 22.94 | 103290.67 |
| **BSP** | 5.19 | 31.47 | 16.65 | 13.79 | 21114.23 |
| **AXF** | 6.57 | 31.10 | 11.56 | 12.28 | 22126.26 |
| **PSP** | 6.68 | 13.32 | — | 10.52 | 18895.52 |
| **MST** | 6.54 | 30.88 | — | 7.51 | 19.90 |
| **VWAP** | 7.40 | 12.89 | — | 9.31 | 3259.47 |

**Figure 7: Comparison between DBToaster and two commercial query engines (in refreshes per second). Both the DBMS (DBX) and stream system (SPY) columns show the cost of full refresh on each update.**

nested aggregate loops over the full domain of that aggregate even when only a small subset of the domain is affected. We can exploit this to range-restrict variables of the nested subquery. In this example, this is done by the predicate $OK = ok$. If the nested subquery is correlated with the outer query on an equality, this range-restriction is propagated to the outer level, significantly reducing the computational cost.

Repeated application of rewriting rules 2, 3, 4 and 1 with the trigger variable optimization, and pushing down selections results in the following materialization decision:

$$\mathcal{M}(\text{Sum}_{CK;1}(\sigma_{OK=ok}(C \bowtie O))) \bowtie ($$
$$(\mathcal{M}(Q_2) + \{\langle\rangle \mapsto qty\}) \bowtie \sigma_{100<qty+\mathcal{M}(Q_2)}(1)$$
$$- \mathcal{M}(Q_2) \bowtie \sigma_{100<\mathcal{M}(Q_2)}(1))$$
where $Q_2 = \text{Sum}_{\langle\rangle;QTY}(\sigma_{OK=ok}LI)$

Apart from the outermost materialization (of $C \bowtie O$), the remaining four materializations in this expression are not only equivalent, but identical to $Q_{O2}$, which is already being maintained. Only one view: $Q_{LI} := \text{Sum}_{CK;1}(\sigma_{OK=ok}(C \bowtie O))$ is materialized. Rewriting the materialization decision produces trigger statement `08`.

Note that this statement requires a loop. We update $Q$ iterating over $\text{dom}_{CK}(\Delta_{+LI}Q) = \text{dom}_{CK}(Q_{LI})$. For this example, the loop never encounters more than one tuple due to the foreign key dependency from $O$ to $C$.

$Q_{LI}$ can be maintained in a manner analogous to that of Example 3, resulting in trigger statements `03`, `05`, and `06`.

## 6. EXPERIMENTAL RESULTS

The DBToaster compiler produces trigger programs as C++ code with views implemented by Boost MultiIndexes, a flexible main-memory collection data structure supporting a variety of secondary index types. Our compiler internals are ongoing research and outside the scope of this paper. We evaluate the experimental performance of DBToaster on Redhat Enterprise Linux with 16 GB of RAM, and an Intel Xeon E5620 2.4 GHz processor (on a single-core).

**Data and Query Workload.** Our workload captures algorithmic order book trading and online business decision support scenarios that involve computing a variety of statistics to guide actions. Figure 6 lists the processing properties of our workload, with SQL code in Appendix A.

The financial queries VWAP, MST, AXF, BSP, PSP, and BSV were run on a 2.63 million tuple trace of an order book update stream, representing one day of stock market activity for MSFT. These are updates to a Bids and Asks table with a schema of a timestamp, an order id, a broker id, a price, and a volume. The TPC-H benchmark queries Q3, Q11, Q17, Q18, Q22, and SSB4 were run on a stream of updates adapted from a database generated by DBGEN[31]. We simulate a system that monitors a set of "active" orders by randomly interleaving insertions on all relations and injecting random deletions on Orders rows to keep the Orders table at around 30 thousand tuples. Results presented in Figures 8, 9 and 10 are based on a scaling factor 0.1 (100 MB) database. We show that these results scale to longer streams in Section 6.2.

To evaluate our compilation algorithm, DBToaster produces three alternatives by terminating recursive compilation early. Depth 0 compilation corresponds to re-evaluating the query on every update, while compilation at Depth 1 is classical first-order IVM. As the third option, DBToaster materializes as much of the query as possible (Naive Recursive), creating view caches and employing partial materialization to decorrelate nested subqueries. Our results show the number of tuples processed by queries run over a replayed stream for a 1 hour period.

### 6.1 Higher-Order IVM Performance

We now analyze the steady-state performance of DBToaster.

**Comparison with Commercial Systems.** Figure 7 compares higher-order IVM to the performance of a commercial DBMS (DBX) and stream processor (SPY), anonymized due to their license agreements. We present a summary of our findings, an in-depth itemized breakdown of overheads is outside the scope of this paper. SPY does not support IVM, so the number presented is for full re-evaluation of the query on every update. For nested queries, our workload utilizes SPY's in-memory tables which significantly contributes to the performance gap from DBToaster due to their synchronization requirements in an asynchronous stream en-
976

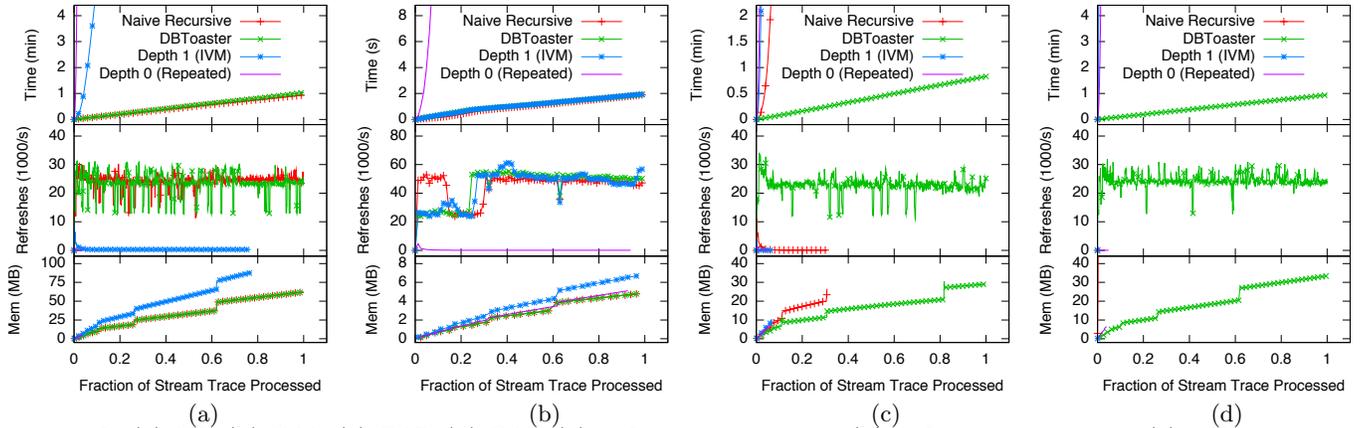

Figure 8: (a) Q3, (b) Q11, (c) Q17, (d) Q18; (a) A 3-way linear join. (b) A 2-way linear join. (c) A 2-way join with an equality-correlated nested aggregate. (d) A 3-way join with an equality-correlated nested aggregate.

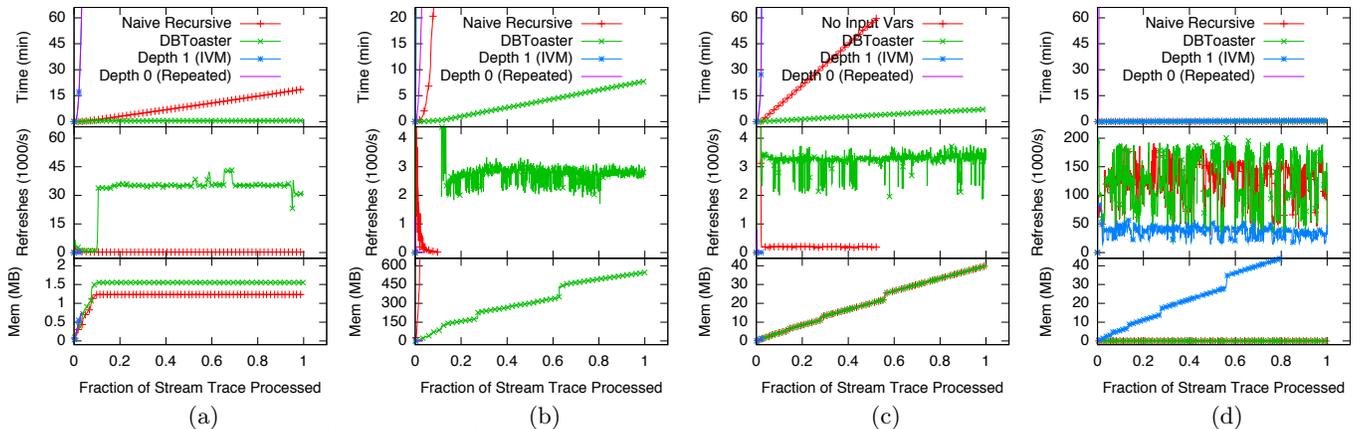

Figure 9: (a) Q22, (b) SSB4, (c) VWAP, (d) BSV; (a) A single table with an equality- and an inequality-correlated nested aggregates. Insertions into the Customer relation complete within the first 10% of the stream. (b) A 3-way star join with a maximum join width of 6. (c) A single table with an inequality-correlated and an uncorrelated nested aggregate. DBToaster chooses the naive recursive approach, so we compare against an approach that aggressively avoids input variables. (d) A 2-way self-join.

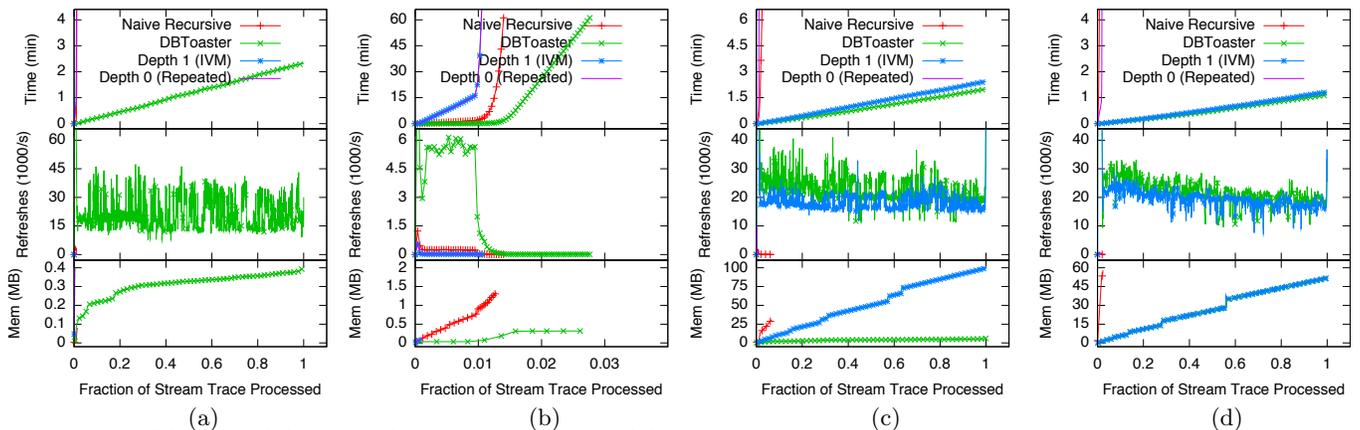

Figure 10: (a) PSP, (b) MST, (c) AXF, (d) BSP; (a) A 2-way join with two uncorrelated nested aggregates. (b) A 2-way join with two uncorrelated, and two inequality-correlated nested aggregates. None of the tested engines completed the trace within the 60 minute limit. (c) A 2-way inequality join. (d) An inequality self-join.

977

gine. For DBX, although this does support IVM, more than half of the queries in our test workload require features of SQL that cannot be handled incrementally by DBX's views subsystem. In our experiments, we found two significant contributors to DBX's overheads. First, because DBX only performs IVM after commits, transaction overheads greatly add to the cost of achieving fast refresh. Second, maintaining catalog information across many tables for high-rate updates also substantially impacts latencies and throughput.

**Equijoins.** Q3 and Q11 (Figure 8a,b) are 2- and 3-way linear joins respectively, SSB4 (Figure 9b) is a 3-way star join with a maximum join-width of 6, and BSV (Figure 9d) is a 2-way self-join. As there are no inequalities, DBToaster and Naive Recursive Compilation produce mostly identical results. In the case of many 2-way joins, the first level deltas are very nearly the base relations, and so on Q11, IVM is able to perform as effectively as DBToaster. On BSV however, DBToaster gets a substantial performance improvement by representing the materialized delta view with only a single aggregate value, making the update cost constant. SSB4 normally has a join width of 6. However, because the contents of the Nation table are static, DBToaster does not attempt to materialize any deltas needed to support updates to Nation, reducing the join width to 4 and eliminating several maps with high maintenance costs.

**Nested Aggregates.** Q17 and Q18 (Figure 8c,d) are multiway join nested-aggregate queries with simple nested aggregates, in both case with the nested aggregate correlated on an equality. Here, DBToaster's strong performance comes from decorrelating the nested subquery for only the deltas of Lineitem (on which both nested subqueries are based).

Q22 (Figure 9a) includes two nested aggregates, an uncorrelated aggregate on Customer that is compared against on the top level using an inequality and an equality-correlated aggregate on Orders that is compared against using an inequality. The first nested subquery causes DBToaster to choose a strategy of re-evaluating the top-level query since the delta of the subquery with respect to updates to Customer is not simpler than the original subquery. The second subquery by itself would not have made this necessary since we can decorrelate it (due to the absence of both inequalities and the Customer relation) and avoid input variables in any query subexpression. Nevertheless, the two subqueries as well as the top-level aggregation without the inequality can be materialized, reducing re-evaluation to a loop over nations. This is seen in the performance graph as the query's slow startup ends once the last customer has been inserted.

VWAP (Figure 9c) has a nested aggregate correlated on an inequality. The small domain of the correlation variable (price) makes this an ideal candidate for view caching.

PSP (Figure 10a) has two uncorrelated nested aggregates. This query benefits from full re-evaluation on each execution. However, polynomial expansion actually enables a graph decomposition that splits the query into 4 parts: 2 constant time components and 2 independent linear time components in the number of distinct values of the column being compared to the nested aggregate (volume).

MST (Figure 10b) is fundamentally similar to PSP, but rather than comparing its uncorrelated aggregates against columns from the base relations, they are each compared against another nested aggregate correlated on an inequality. This is a worst case scenario for DBToaster, as it cannot incrementally process this query in better than $O(n^2)$ time without specialized indexes (e.g., aggregate range trees).

**Inequijoins.** AXF and BSP are both 2-way joins (Figure 10c,d), with BSP being a self-join. In the case of AXF, both the join variable (price) and one of the aggregate variables (volume) are treated as input variables by naive recursive materialization. In BSP, the join variable (t) has an extremely large domain. In both cases, partial materialization outperforms naive. Since both are 2-way joins, IVM is nearly optimal – DBToaster achieves a small speed boost in both cases by not materializing the entire base relation.

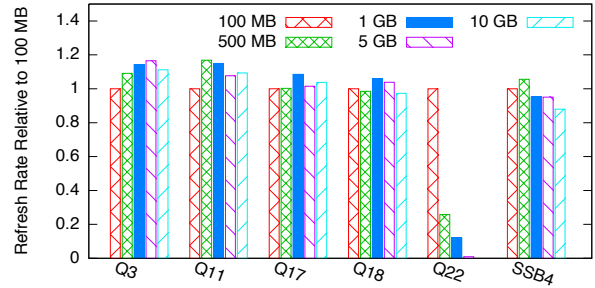

**Figure 11: Performance scaling on the TPCH queries.**

## 6.2 Working State Scalability

Figure 11 analyzes the performance scaling properties of DBToaster on larger datasets and subsequently a larger working state for its main-memory data structures. An update stream was synthesized from databases created by DBGEN at scaling factors 0.5, 1, 5, and 10 (500MB, 1GB, 5GB, and 10GB respectively). As before, the Orders relation is kept at 30 thousand tuples. The Customer, Part, Supplier, or Partsupp are inserted completely and never deleted. With the exception of Q22, performance stays roughly constant as stream length grows.

The running time of Q22 is dominated by the first 10% of the stream in each case, before the customers table has been fully inserted. The cost of each insertion into the customer table is linear in the size of the customer table. After all customer tuples have been loaded in, performance returns to a constant 35000 tuples per second, regardless of scale.

## 7. CONCLUSION

We presented a compiler and optimizer framework for higher-order IVM that uses aggressive simplification of recursive delta queries and a plethora of materialization strategies to make recursive IVM viable. Our compilation method is effective on a wide range of select-project-join-aggregate queries, including those with nested subqueries that are not supported by current IVM mechanisms. Our methods provide scalable view refresh rates, often orders of magnitude over today's tools, providing the basis for our vision of DDMS.

## Appendix A.    QUERIES

AXF
```
SELECT   b.broker_id, sum(a.volume-b.volume)
FROM     Bids b, Asks a
WHERE    b.broker_id = a.broker_id
  AND    (a.price-b.price > 1000 OR b.price-a.price > 1000)
GROUP BY b.broker_id;
```

BSP
```
SELECT x.broker_id, sum(x.volume*x.price - y.volume*y.price)
FROM   Bids x, Bids y
WHERE  x.broker_id=y.broker_id AND x.t>y.t
GROUP BY x.broker_id
```

BSV
```
SELECT x.broker_id, sum(x.volume*x.price*y.volume*y.price*0.5)
FROM   Bids x, Bids y
WHERE  x.broker_id = y.broker_id GROUP BY x.broker_id;
```

MST
```
SELECT b.broker_id, sum(a.price*a.volume - b.price*b.volume)
FROM   Bids b, Asks a
WHERE 0.25*(select sum(a1.volume) from Asks a1) >
   (select sum(a2.volume) from Asks a2 where a2.price>a.price)
  AND   0.25*(select sum(b1.volume) from Bids b1) >
   (select sum(b2.volume) from Bids b2 where b2.price>b.price)
GROUP BY b.broker_id;
```

PSP
```
SELECT sum(a.price - b.price) FROM Bids b, Asks a
WHERE  b.volume>0.0001*(select sum(b1.volume) from Bids b1)
AND    a.volume>0.0001*(select sum(a1.volume) from Asks a1);
```

VWAP
```
SELECT sum(b1.price * b1.volume) FROM Bids b1
WHERE  0.25 * (select sum(b3.volume) from Bids b3) >
       (select sum(b2.volume) from Bids b2
          where b2.price>b1.price);
```

Q3
```
SELECT o.orderkey, o.orderdate, o.shippriority,
       sum(li.extendedprice * (1 - li.discount))
FROM   Customer c, Orders o, Lineitem li
WHERE  c.mktsegment = 'BUILDING'
  AND  o.custkey = c.custkey
  AND  li.orderkey = o.orderkey
  AND  o.orderdate < DATE('1995-03-15')
  AND  li.shipdate > DATE('1995-03-15')
GROUP BY o.orderkey, o.orderdate, o.shippriority;
```

Q11
```
SELECT ps.partkey, sum(ps.supplycost * ps.availqty)
FROM   Partsupp ps, Supplier s
WHERE  ps.suppkey = s.suppkey GROUP BY ps.partkey;
```

Q17
```
SELECT sum(l.extendedprice) FROM Lineitem l, Part p
WHERE  p.partkey = l.partkey
AND    l.quantity < 0.005 * (select sum(l2.quantity)
          from Lineitem l2 where l2.partkey = p.partkey);
```

Q18
```
SELECT c.custkey, sum(l1.quantity)
FROM Customer c, Orders o, Lineitem l1
WHERE 1 <= (select sum(1) where
            100 < (select sum(l2.quantity) from Lineitem l2
                   where l1.orderkey = l2.orderkey))
AND c.custkey = o.custkey AND o.orderkey = l1.orderkey
GROUP BY c.custkey;
```

Q22
```
SELECT c1.nationkey, sum(c1.acctbal) FROM Customer c1
WHERE c1.acctbal <
   (select sum(c2.acctbal) from Customer c2 where c2.acctbal>0)
AND 0=(select sum(1) from Orders o where o.custkey=c1.custkey)
GROUP BY c1.nationkey;
```

SSB4
```
SELECT sn.regionkey, cn.regionkey, p.type,
       sum(li.quantity)
FROM   Customer c, Orders o, Lineitem li,
       Part p, Supplier s, Nation cn, Nation sn
WHERE  c.custkey = o.custkey
  AND  o.orderkey = li.orderkey
  AND  p.partkey = li.partkey
  AND  s.suppkey = li.suppkey
  AND  o.orderdate >= DATE('1997-01-01')
  AND  o.orderdate <  DATE('1998-01-01')
  AND  cn.nationkey = c.nationkey
  AND  sn.nationkey = s.nationkey
GROUP BY sn.regionkey, cn.regionkey, p.type;
```